\documentclass[prb,twocolumn,showpacs,floatfix]{revtex4-1}
\usepackage{graphics,epsfig,amsfonts,amssymb,amsmath,soul}
\begin{document}

\title{Head-to-Head Domain Wall Structures in Wide Permalloy Strips}

\author{Virginia Est\'evez}
\email{virginia.esteveznuno@aalto.fi}
\author{Lasse Laurson}

\affiliation{COMP Centre of Excellence and Helsinki Institute of Physics,
Department of Applied Physics, Aalto University, P.O.Box 11100, 
FI-00076 Aalto, Espoo, Finland.}

\begin{abstract}
We analyze the equilibrium micromagnetic domain wall structures 
encountered in Permalloy strips of a wide range of thicknesses and widths,
with strip widths up to several micrometers. By performing an extensive
set of micromagnetic simulations, we show that the equilibrium phase 
diagram of the domain wall structures exhibits in addition to the previously 
found structures (symmetric and asymmetric transverse walls, vortex wall) 
also double vortex and triple vortex domain walls for large enough strip 
widths and thicknesses. Also several metastable domain wall 
structures are found for wide and/or thick strips. We discuss 
the details of the relaxation process from random magnetization initial 
states towards the stable domain wall structure, and show that our results
are robust with respect to changes of e.g. the magnitude of the Gilbert 
damping constant and details of the initial conditions.
\end{abstract}
\maketitle

\section{Introduction}

During the last decade, a lot of effort has been devoted  
to understand static and dynamic properties of magnetic domain walls 
(DWs) in ferromagnetic nanostructures such as nanowires and -strips. 
These studies have been largely driven by promising technological 
applications based on domain walls and their dynamics, in particular 
memory \cite{barnes2006, parkinracetrack} and logic 
devices \cite{cowburn2000,allwood2002,allwood2005}. In typical experiments 
DWs are driven by either applied magnetic fields \cite{ono99,beach2005} 
or spin-polarized electric currents \cite{spcklaui2003,thiaville2004,klaui2005,
vernier2004}. The resulting DW dynamics
depends crucially on the micromagnetic DW structure, typically involving 
various internal degrees of freedom. These are essential e.g. for the 
emergence of the Walker breakdown \cite{walker74}, an instability occurring
when the DW internal degrees of freedom get excited by 
a strong external drive (a magnetic field $H$ or a spin-polarized
current $J$ exceeding the Walker field $H_\text{W}$ or current $J_\text{W}$, 
respectively), limiting the propagation velocity of the DWs.

Two main classes of ferromagnetic materials have been extensively studied 
within the strip geometry. Materials with a high perpendicular 
magnetic anisotropy (PMA \cite{buschowbook,coeybook,boulle2008, martinez2012}) 
exhibit simple and narrow DWs of the Bloch and/or N\'eel type. For
$H > H_\text{W}$ or $J > J_\text{W}$, 
repeated transitions between these two structures are observed \cite{martinez2012}. 
The second class of systems includes soft (low anisotropy) magnetic 
materials \cite{buschowbook,coeybook} such as Permalloy, where in-plane 
domain magnetization along the long axis of the strip is induced by shape 
anisotropy. By using various experimental techniques \cite{parkinhandbook,klauireview} 
and micromagnetic simulations, it has been established 
that the equilibrium DW structures separating these in-plane domains are more
complex, and depend crucially on the sample geometry
\cite{reviewthiaville,McMichael97,nakatani2005,klaui2004}. 
Transverse DWs (TWs) and asymmetric transverse DWs (ATWs) are observed for 
narrow and thin strips \cite{prbklaui2003,JAM-14,nozaki99,klaui2004,annularklaui2006,atwbackes2007}, 
while in wider and thicker strips one encounters the vortex 
DW (VW) \cite{nozaki99,klaui2004,prbklaui2003,vortexklaui2006,annularklaui2006,hempe2007}. 
In addition, various metastable DW structures with higher energy may be
found \cite{klauireview,vortexklaui2006,hempe2007,vaz2005,park2006}.
For $H > H_\text{W}$ or $J > J_\text{W}$, the DW structures exhibit dynamical
evolution: for TWs, repeated nucleation and propagation of an antivortex 
across the strip width takes place \cite{reviewthiaville}. Similarly, 
in VWs the vortex core performs oscillatory back and forth perpendicular 
motion \cite{reviewthiaville}. 

In Permalloy strips with even larger widths and/or thicknesses, one might
expect also other, possibly more complicated equilibrium DW 
structures. For wider strips shape anisotropy is less important, implying 
that energy minima with more complex spin structures closing the flux more 
efficiently than TWs, ATWs or VWs may appear. Indeed, e.g. double and 
triple vortex DWs have been observed in experiments on wide strips \cite{vortexklaui2006}, 
but they have been attributed to current-induced vortex nucleation resulting in 
metastable DW structures.  
Consequently, a pertinent and fundamental question is what are the possible 
intermediate equilibrium DW structures observable when the lateral 
Permalloy strip dimensions increase from those corresponding to the typical 
nanostrip geometry (with TW, ATW or VW as the stable DW structure) to strip 
widths of micrometers and beyond.

In this paper we present an extensive numerical study of the equilibrium and 
metastable micromagnetic DW structures in Permalloy strips, with the strip
widths up to an order of magnitude larger than before \cite{reviewthiaville,McMichael97,
nakatani2005,klaui2004,JAM-14,nozaki99}. Contrary to previous studies focusing on 
comparing the energies of different {\it a priori} known DW structures \cite{reviewthiaville,
McMichael97,nakatani2005,klaui2004,JAM-14,nozaki99}, we perform micromagnetic 
simulations of relaxation dynamics from random initial states towards the 
stable DW structures. In addition to the previously observed TW, ATW and VW DWs, 
we find also DWs with equilibrium double and triple vortex structures for wide and/or 
thick enough strips. The last structure is encountered only in the very largest system 
sizes we were able to simulate. Moreover, for wide strips we find a rich variety 
of metastable DWs with even more complex micromagnetic structures. We demonstrate
that our results are robust with respect to changes of the magnitude of 
the Gilbert damping constant or using different initial conditions for the 
relaxation process. Our results underline the crucial role of topological 
defects for physics of DWs in soft strips, and that of micromagnetic
simulations for finding the true equilibrium DW structure.

\begin{figure}[t!]
\leavevmode
\includegraphics[trim=1.5cm 5.2cm 3cm 4.5cm, clip=true,width=0.9\columnwidth]{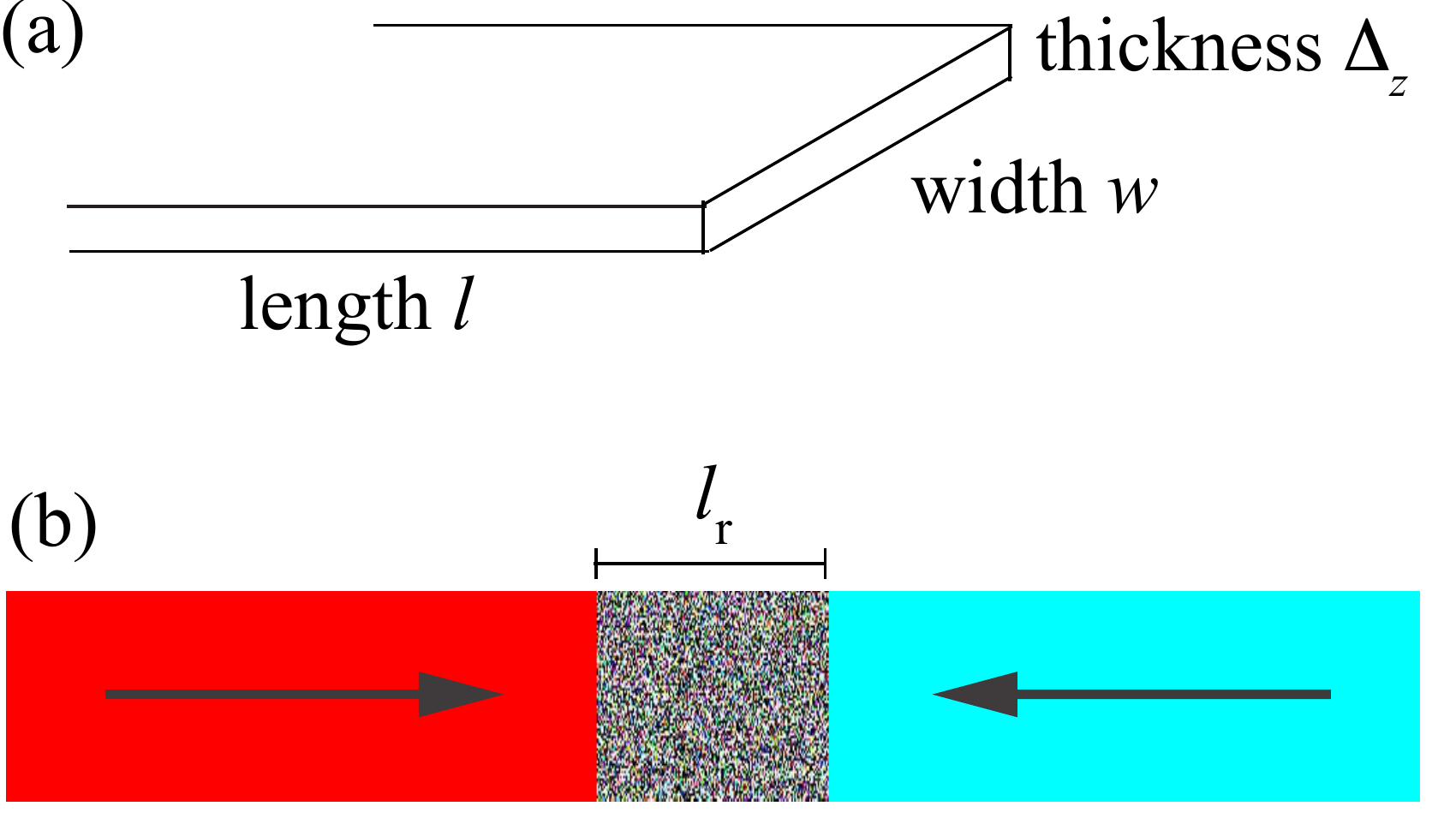}
\caption{(color online) (a) Geometry of the Permalloy strip. 
(b) A top view of the magnetization in the initial state. Magnetization points 
along the long axis of the strip within the
two domains (as indicated by the arrows) forming a head-to-head configuration. 
In between them, a region of random magnetization (of length $l_\text{r}$)
has been included.}
\label{fig:fig1_equilibriumphase}
\end{figure}

\section{Micromagnetic simulations}

The system studied is a Permalloy 
strip of width $w$ and thickness $\Delta_z$, satisfying
$\Delta_z \ll w$, see Fig.~\ref{fig:fig1_equilibriumphase} (a). In the micromagnetic
simulations, magnetic charges are compensated on the left and right ends of
the strip, to mimic an infinitely long strip; the actual simulated length
satisfies $l \geq 4w$ for all cases considered. The initial state
from which the relaxation towards a stable DW structure starts is
an in-plane head-to-head domain structure, with a region of random magnetization
of length $l_\text{r}$ in the middle of the sample, see 
Fig.~\ref{fig:fig1_equilibriumphase} (b). If not specified otherwise, 
we consider $l_\text{r} = 2w$. Material parameters of Permalloy
are used, i.e. saturation magnetization $M_\text{s}=860 \times 10^3$ A/m and exchange
constant $A_\text{ex} = 13 \times 10^{-12}$ J/m. The typical Gilbert damping
constant for Permalloy is $\alpha=0.01$, but here we analyze also the
influence of $\alpha$ on the relaxation process, and thus consider also other
values. For simplicity, we set the temperature $T$ to zero, and focus on the ideal 
case of strips free of any structural disorder or impurities.

The simulations are performed using the GPU-accelerated micromagnetic code 
MuMax3 \cite{mumax3,mumax2011,mumax2014}, offering a significant speedup
as compared to CPU codes for the large system sizes we consider here.
To calculate the magnetization dynamics of the system, the Landau-Lifshitz-Gilbert 
equation \cite{gilbert2004,brownmicromagnetism},
\begin{equation}
\partial {\bf m}/\partial t =
\gamma {\bf H_{eff}} \times {\bf m} + \alpha {\bf m} \times
\partial {\bf m}/\partial t,
\end{equation}
is solved numerically. Here, ${\bf m}$ is the magnetization, $\gamma$ the 
gyromagnetic ratio, and ${\bf H}_\text{eff}$ the effective
field, with contributions due to exchange, Zeeman, and demagnetizing
energies. The size of the discretization cell used depends on the system 
size, but is always bounded by the exchange length, $\Lambda = (2A/\mu_0 M_s^2)^{1/2}
\approx 5$ nm, in the in-plane directions, and equals $\Delta_z$ in the the out-of-plane
direction. 

\begin{figure}[t!]
\leavevmode
\includegraphics[trim=1.5cm 9.5cm 1.5cm 4.3cm, clip=true,width=0.9\columnwidth]{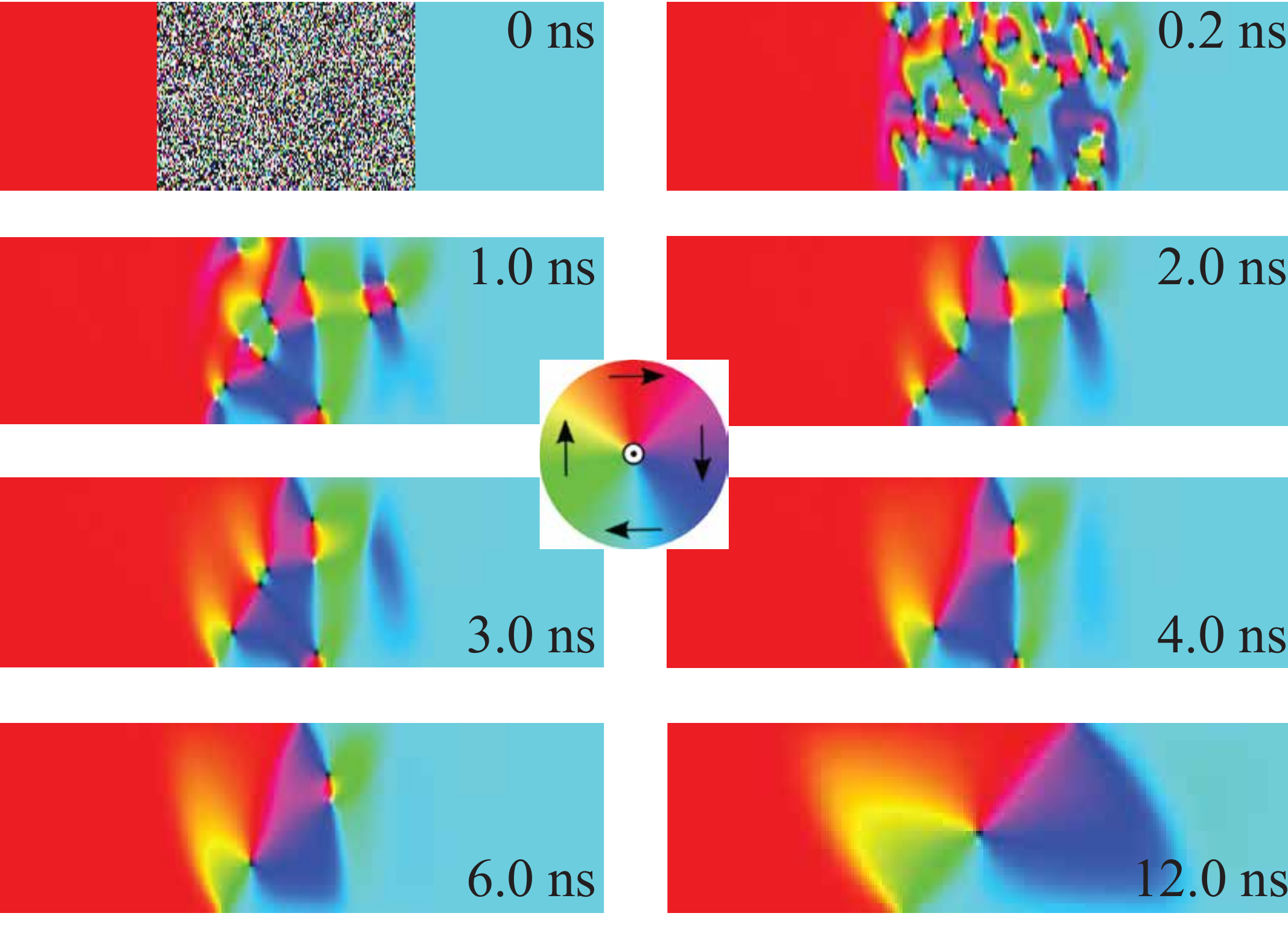}
\caption{(color online) An example of the temporal evolution of the relaxation, 
with $w=420$ nm, $\Delta_z=10$ nm, $\alpha=3$ and $l_\text{r}=2w$. 
Relaxation towards the equilibrum DW structure (here, a VW) 
takes place via coarsening dynamics of the defect structure in the magnetic 
texture. The colorwheel in the middle shows the mapping between magnetization 
directions and colors.}
\label{fig:fig2_equilibriumphase}
\end{figure}

\section{Results}

\begin{figure}[t!]
\leavevmode
\includegraphics[clip,width=0.9\columnwidth]{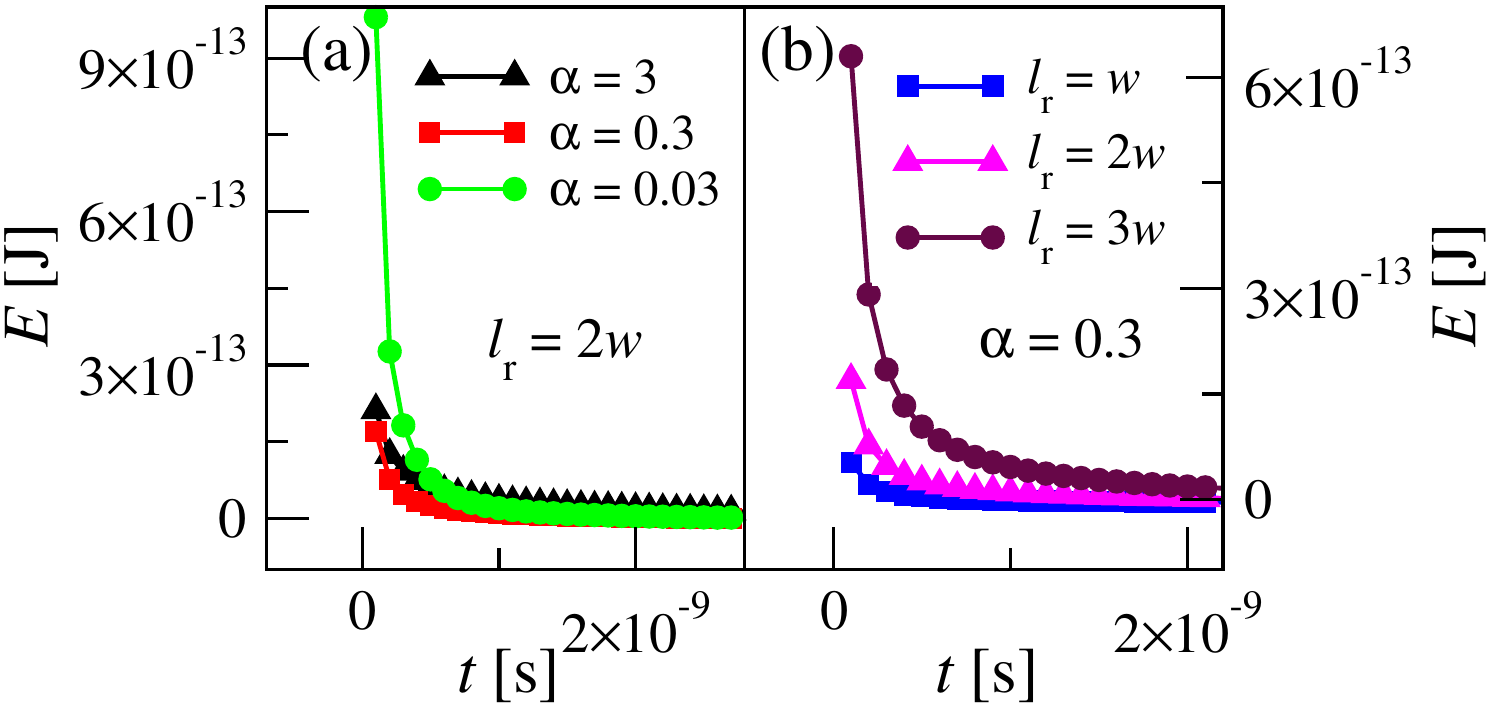}\\
\hspace{-0.6cm}
\includegraphics[trim=0cm 0cm 0cm 0cm, clip=true,width=0.8\columnwidth]{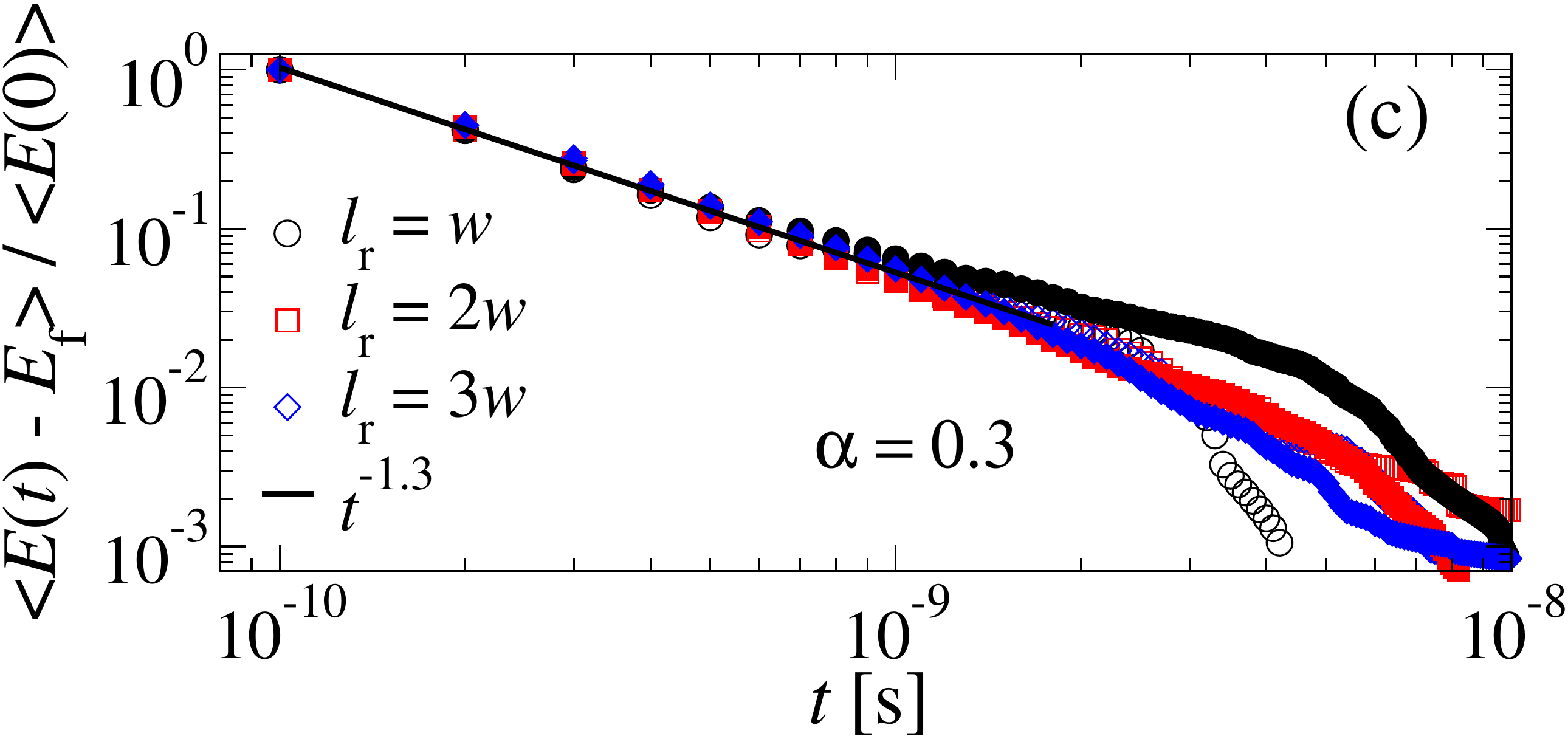}
\caption{(color online) The energy $E(t)$ as a function of time $t$ for 
$w = 5120$ nm and $\Delta_z = 20$ nm. (a) For different values of $\alpha$  
and $l_\text{r} = 2w$. (b) For different values of $l_\text{r}$, and $\alpha=0.3$ 
[resulting in the fastest relaxation in (a)]. (c) shows that on average,
the early time decay of $E(t)$ towards its final value $E_\text{f}$ obeys 
$\langle E(t)-E_\text{f} \rangle \propto t^{-\beta}$. For the $\alpha=0.3$ case 
shown here, $\beta \approx 1.3$. Empty (filled) symbols in (c) correspond to 
$w=420$, $\Delta_z=10$ nm ($w=860$, $\Delta_z=20$ nm).}
\label{fig:fig3_equilibriumphase}
\end{figure}

We start by considering the effect of varying $\alpha$ and
$l_\text{r}$ on the relaxation process.
Fig.~\ref{fig:fig2_equilibriumphase} shows an example of the time evolution of 
${\bf m} ({\bf r},t)$ for $w = 420$ nm, $\Delta_z = 10$ nm, $\alpha = 3$ and $l_\text{r} = 2w$. 
The initially random magnetization evolves via coarsening of the defect structure 
of the magnetization texture towards the stable DW (here, a VW).
During the relaxation, the total energy $E$ of the system decreases in a manner
that for a given geometry ($w$ and $\Delta_z$) depends on both $\alpha$ and $l_\text{r}$,
see Fig.~\ref{fig:fig3_equilibriumphase} (a) and (b) where $w = 5120$ nm and 
$\Delta_z = 20$ nm, is considered. For instance, $E$ decreases faster 
for an intermediate $\alpha$ [Fig.~\ref{fig:fig3_equilibriumphase} (a)]. 
We attribute this behavior to the balance between inertial effects related
to precession favored by a small $\alpha$, helping to overcome energy barriers, 
and the higher rate of energy dissipation due to a large $\alpha$. Thus, the 
relaxation time to reach a (meta)stable DW structure depends on 
$\alpha$. Fig.~\ref{fig:fig3_equilibriumphase} (b) illustrates that 
for a fixed $\alpha$, systems with a larger $l_\text{r}$ relax more slowly.
Fig.~\ref{fig:fig3_equilibriumphase} (c) shows that on average, the early-time
relaxation of $E$ towards its final value $E_\text{f}$ exhibits temporal power-law 
decay, $\langle E-E_\text{f} \rangle \propto t^{-\beta}$ with $\beta \approx 1.3$ 
for the $\alpha=0.3$ case shown, possibly related to collective effects 
due to interactions between several topological defects during early 
stages of relaxation (Fig. \ref{fig:fig2_equilibriumphase}).

In general, the final (meta)stable DW structure may depend on the realization 
of the random initial state. Thus, we consider 21 realizations of the initial random
magnetization for each $w$ and $\Delta_z$, and compare the energies of the resulting 
relaxed configurations. The structure with the lowest energy is chosen as 
the equilibrium structure, while others with higher energy are metastable 
states. Although, as discussed above, the relaxation times depend on $\alpha$
and $l_\text{r}$, the equilibrium DW structure is found to be independent of 
$\alpha$ and $l_\text{r}$ in the range considered, i.e. $\alpha \in 
[0.01, 3]$ and $l_\text{r} \in [w, 3w]$. Thus, in what follows, we will 
use $\alpha = 3$ and $l_\text{r} = 2w$. 

\begin{figure}[t!]
\leavevmode
\includegraphics[trim=0.3cm 7.75cm 1.25cm 8.5cm, clip=true, 
width=0.9\columnwidth]{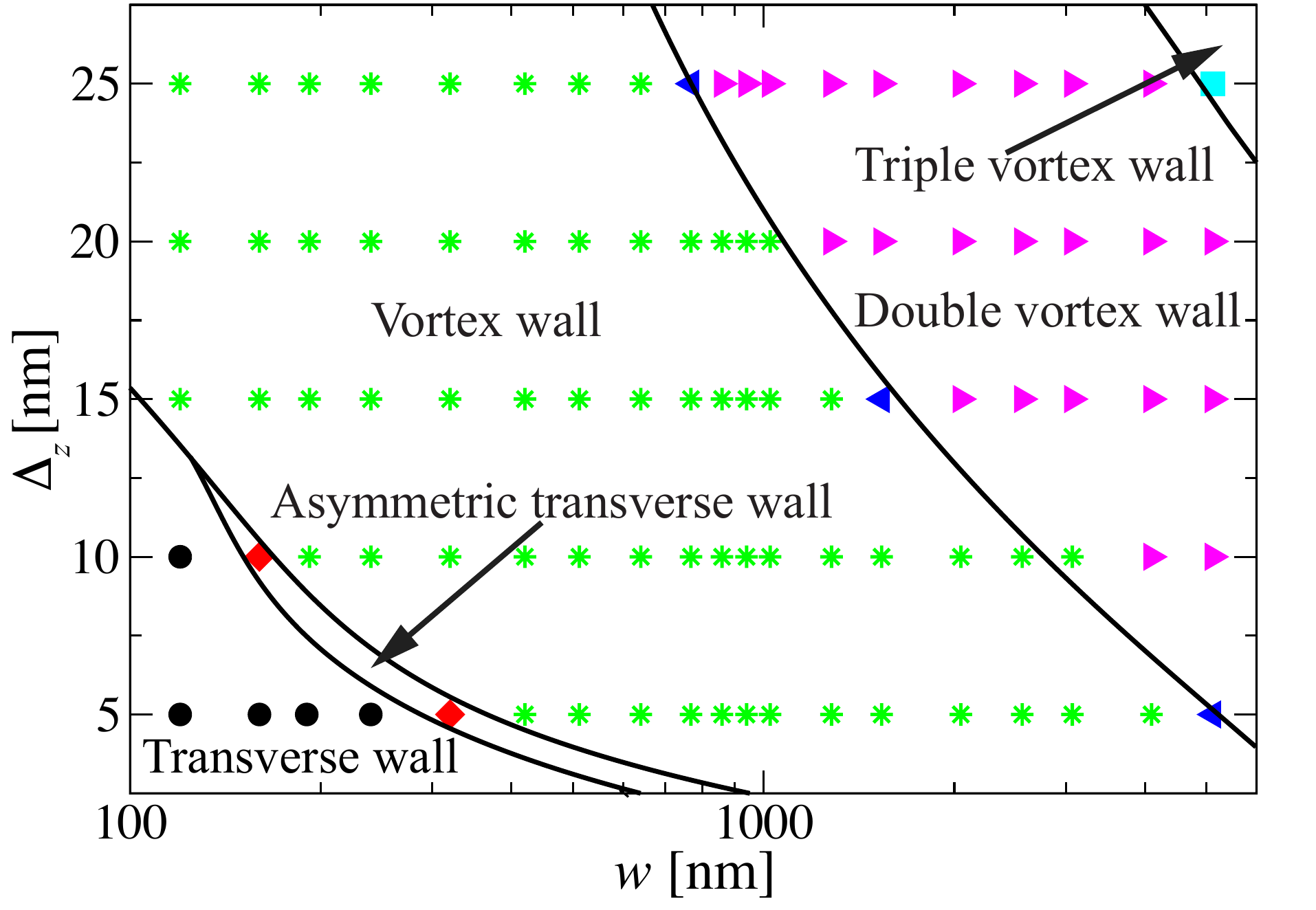}
\caption{(color online) Phase diagram of the equilibrium DW structure 
in Permalloy strips of various thicknesses (from $\Delta_z = 5$ to 25 nm) and widths ranging 
from $w = 120$ nm up to $5120$ nm. The symbols correspond to observations
of the various equilibrium DW structures, with phase boundaries shown as solid
lines. Examples of the DW structures corresponding to the 5 different phases 
are shown in Fig. \ref{fig:fig5_equilibriumphase}.}
\label{fig:fig4_equilibriumphase}
\end{figure}

\begin{figure}[t!]
\leavevmode
\includegraphics[trim=0.25cm 2.25cm 0.25cm 2.0cm, clip=true,width=0.9\columnwidth]{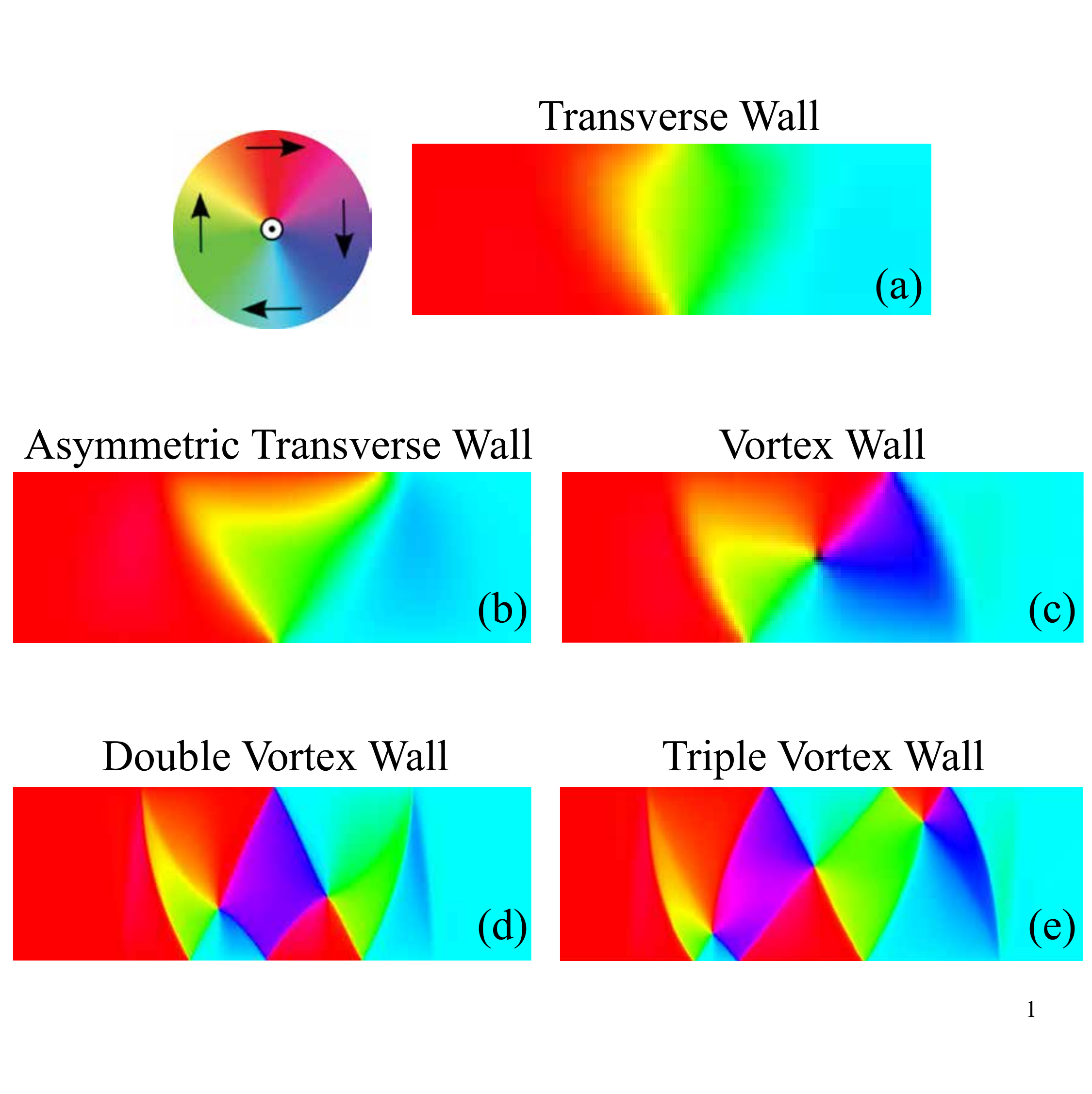}
\caption{(color online) Examples of the different equilibrium micromagnetic 
DW structures: (a) TW for $w = 120$ nm and $\Delta_z = 5$ nm, (b) ATW for $w = 160$ 
nm and $\Delta_z = 10$ nm, (c) VW for $w = 640$ nm and $\Delta_z = 15$ nm, (d) DVW for 
$w = 2560$ nm and $\Delta_z = 20$ nm, and (e) TVW for $w = 5120$ nm and $\Delta_z = 25$ nm. 
The colorwheel (top left) shows the mapping between magnetization directions 
and colors.}

\label{fig:fig5_equilibriumphase}
\end{figure}

The main results of this paper are summarized in Figs.~\ref{fig:fig4_equilibriumphase}
and \ref{fig:fig5_equilibriumphase}, showing the phase diagram of the equilibrium
DW structures for $w$ ranging from $120$ to $5120$ nm, and $\Delta_z$ from $5$ to $25$ nm, 
and examples of these structures, respectively. For small
$w$, we recover the previous results \cite{reviewthiaville,McMichael97,
nakatani2005,klaui2004}, i.e. phases correspoding to TW, ATW and VW, 
shown in Fig.~\ref{fig:fig5_equilibriumphase} (a), (b) and (c), respectively.

For larger strip widths ($w$ approaching or exceeding $1 \mu$m, depending on $\Delta_z$,
see Fig.~\ref{fig:fig4_equilibriumphase}), a new equilibrium micromagnetic DW 
structure, a double vortex wall (DVW), is observed. This structure consists of 
two vortices with opposite sense of rotation of the magnetization around the vortex 
core, see Fig.~\ref{fig:fig5_equilibriumphase} (d). At the phase boundary
(blue triangle symbols pointing left in Fig.~\ref{fig:fig4_equilibriumphase}), 
VW and DVW have the same energy. The DVW phase spans a relatively large area within 
the $(w,\Delta_z)$ space, highlighting the robustness of our results. 

In addition, a second new phase, with a triple-vortex wall (TVW) as the equilibrium
structure [see Fig.~\ref{fig:fig5_equilibriumphase} (e)], is found for the very 
largest system sizes we have been able to simulate. The middle vortex of the TVW has
an opposite sense of rotation to the other two. For $w=5120$ and $\Delta_z=25$ nm, 
DVW and TVW have the same energy (the cyan square symbol in the top right corner of 
Fig.~\ref{fig:fig4_equilibriumphase}), suggesting the presence of 
a phase boundary between the two structures. Indeed, by performing a set of 10 
additional simulations with $w=6144$ and $\Delta_z=25$ nm (i.e. outside the phase 
diagram in Fig.~\ref{fig:fig4_equilibriumphase}), suggests that TVW is the equilibrium 
DW structure for very large strip widths. This structure has been observed in 
experiments as a metastable state for smaller systems \cite{vortexklaui2006,
hempe2007}. Notice also that the middle part of the TVW 
[Fig.~\ref{fig:fig5_equilibriumphase} (e)], exhibiting four line-like
90$^{\circ}$ DWs meeting at a vortex core in the middle of the TVW, 
resembles the typical Landau flux-closure magnetization patterns observed for 
rectangular Permalloy thin films \cite{SHI-00,WAC-02,FIS-11}. 

\begin{figure}[t!]
\leavevmode
\includegraphics[trim=0.25cm 5cm 0.25cm 1.7cm, clip=true, width=0.9\columnwidth]{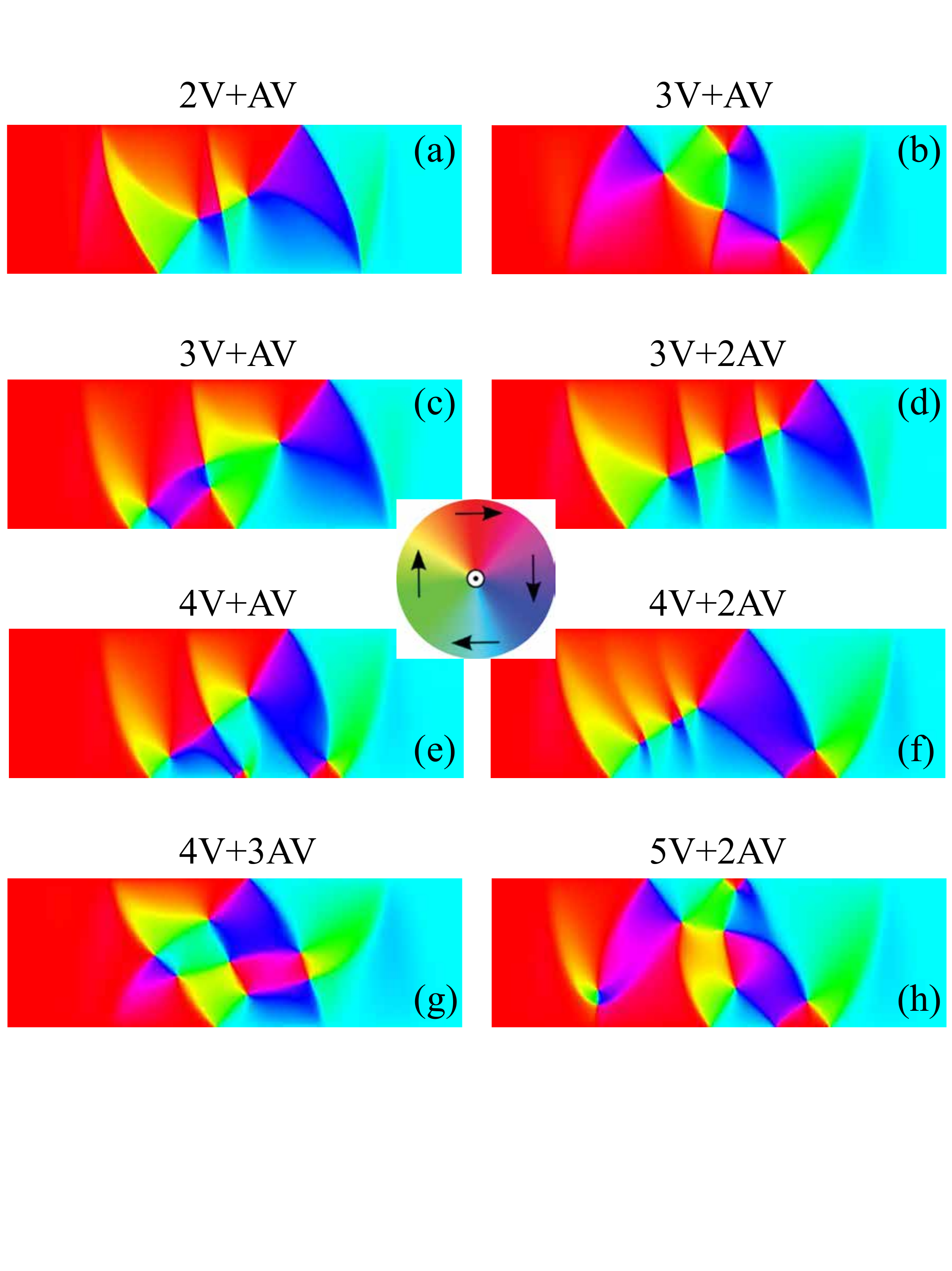}
\caption{(color online) Examples of metastable DW structures observed
for a system with $w = 5120$ nm and different thicknesses $\Delta_z$:
(a) Two vortices and an antivortex (2V+AV), $\Delta_z = 25$ nm, (b) and (c) three 
vortices and an antivortex (3V+AV), $\Delta_z = 5$ nm, (d) three vortices and 
two antivortices (3V+2AV), $\Delta_z = 10$ nm, (e) four vortices and an 
antivortex (4V+AV), $\Delta_z = 5$ nm, (g) four vortices and three antivortices 
(4V+3AV), $\Delta_z = 5$ nm, and (h) five vortices and two antivortices (5V+2AV), 
$\Delta_z = 5$ nm.}
\label{fig:fig6_equilibriumphase}
\end{figure}

Following the relaxation from a random magnetization initial state, the 
system may in general end up
into various metastable states with higher energy than that of the 
equilibrium DW. Sometimes these metastable states have even a 
higher probability than the equilibrium one, estimated here from the 
sample of 21 relaxed configurations. 
Fig.~\ref{fig:fig6_equilibriumphase} shows some of the metastable states 
found for a large strip with $w=5120$ nm and different values of $\Delta_z$;
for strips with smaller lateral dimensions, different metastable states tend to
be less numerous and have a simpler structure.
Despite their apparent complexity, all the metastable DW structures shown 
in Fig.~\ref{fig:fig6_equilibriumphase} respect the basic principles
of topology of DWs. Each of the DWs are composed of topological defects,
with an associated winding number: +1 for vortices, -1 for antivortices,
and $\pm 1/2$ for edge defects \cite{tchernyshyov2005}. In a DW all the 
topological defects have to be compensated, i.e. the total winding number 
is equal to zero. In the case of the DVW, the two 
topological vortex defects are compensated by four edge defects 
[Fig.~\ref{fig:fig5_equilibriumphase} (d)]. For the metastable state 
of two vortices (with the same sense of rotation) and an antivortex [2V+AV,
see Fig.~\ref{fig:fig6_equilibriumphase} (a)], there are two vortices 
and only two edge defects. Thus, in order to compensate the topological 
defects, also an antivortex appears. In general, we have observed that 
in a DW with $N$ vortices with the same sense of rotation, there must 
be $N-1$ antivortices to get a zero total winding number, see 
Fig.~\ref{fig:fig6_equilibriumphase} (a) and (d) for examples with 
2V+AV and 3V+2AV configurations, respectively. When some of the vortices
have oppposite sense of rotation, more complex
scenarios are encountered, with examples shown in 
Figs.~\ref{fig:fig6_equilibriumphase} (b), (c), (e), (f), (g) and (h).
Notice also that two DW structures with the same elements can look very
different, see e.g. the two 3V+AV DWs shown in 
Figs.~\ref{fig:fig6_equilibriumphase} (b) and (c).
All the DW structures found, both the equilibrium ones in 
Fig.~\ref{fig:fig5_equilibriumphase} and the metastable states in
Fig.~\ref{fig:fig6_equilibriumphase} obey the principle of compensation
of topological defects to yield a total winding number of zero. The 
richness of the equilibrium phase diagram and the large collection of 
metastable states indicate that for wide/thick strips in particular, 
the micromagnetic energy landscape is quite complex, with a large number 
of local minima. This is also in agreement with our observations of 
power-law energy relaxation. 

\section{Summary and conclusions}

To summarize, we have performed an extensive set of micromagnetic 
simulations to study the equilibrium and metastable DW structures
in Permalloy strips of a wide range of widths and thicknesses, as well
as the relaxation dynamics starting from random magnetization
initial states. The general trend of our results is that both the 
equilibrium and metastable DW configurations become increasingly complex 
(i.e. they consist of an increasing number of topological defects) as
the lateral strip dimensions increase. We note that somewhat analogous
behaviour - i.e. existence of equilibrium magnetization configurations
with increasing complexity as the system size increases - is observed
also in some other systems such as three-dimensional cylindrical
elements with perpendicular anisotropy \cite{MOU-07,VUK-08}.

Several remarks are in order: first, for strips with even larger lateral 
dimensions one may in principle expect more complex DW patterns - 
possibly with four or more vortices with alternating sense of rotation.
These, however, are currently beyond the reach of our available computing 
resources. Second, our phase diagram allows one to check if experimental 
observations of the various DW structures in wider strips are equilibrium 
configurations or metastable states. According to our review of the 
experimental literature, most observations of DVWs and TVWs appear to be 
metastable states \cite{vortexklaui2006,hempe2007}.
Third, while the equilibrium structures we find are certainly stable in 
the absence of external perturbations such as applied magnetic fields, 
it remains to be seen how their field driven dynamics is like, and 
whether wide strips with a relatively weak shape ansitropy are able to 
support the DWs as compact objects also when external perturbations are 
being applied \cite{ZIN-11}.

\begin{acknowledgments}
We thank Mikko J. Alava for a critical reading of the manuscript.
This work has been supported by the Academy of Finland through its Centres
of Excellence Programme (2012-2017) under project no. 251748, and an Academy
Research Fellowship (LL, project no. 268302). We acknowledge the computational 
resources provided by the Aalto University School of Science ``Science-IT'' 
project, as well as those provided by CSC (Finland).
\end{acknowledgments}

\end{document}